\documentstyle[preprint,revtex]{aps}
\begin{document}
\preprint{WIS--93/121/Dec.--PH}
\draft
\begin{title}
Conductance of Aharonov--Bohm Rings:\\
{}From the Discrete to the Continuous Spectrum Limit.
\end{title}
\author{Alex Kamenev$^1$, Bertrand Reulet$^2$, H\'el\`ene Bouchiat$^2$
\centerline{and Yuval Gefen$^{1,2}$}}
\begin{instit}
$^1$Department of Condensed Matter Physics, The Weizmann Institute of Science,
Rehovot 76100, Israel.\\
$^2$Laboratoire de Physique des Solides, Associ\'e au CNRS, Bat 510,
Universit\'e  Paris--Sud, 91450, Orsay, France.
\end{instit}
\begin{abstract}
The dissipative conductance of an array of mesoscopic rings, subject to an
a.c. magnetic flux is investigated. The magneto--conductance may change sign
between canonical  and grand-canonical statistical ensembles, as function of
the
inelastic level broadening and as function of the temperature. Differences
between canonical  and grand-canonical ensembles persist up to temperature
of the order of the Thouless energy.
\end{abstract}



Many of the idiosyncrasies of mesoscopic systems may be found in the
response of small metallic rings to an a.c. bias. Previous works in
this field have stressed the role of intralevel (diagonal) and
interlevel (off--diagonal) transitions
\cite{Landauer85,Imry86,Trivedi88} as well as the fact that the
overall physics may depend on the type of external bias
\cite{Landauer92} (coupling to external reservoirs, externally
applied electromagnetic fields, {\em etc.}). Several approaches have
been taken to derive quantitatively the average multichannel
conductance of disordered mesoscopic conductors (in the diffusive
regime). We mention here two. One is based on impurity Green's
function techniques. In the context of Aharonov-Bohm cylinders, the
leading order quantum corrections to the Drude conductance has been
first calculated in Ref.\ \cite{Altshuler83}.  Among other
interesting results of this theory, it also predicts small positive
 magneto-conductance (MC), in conjunction with other weak
localization studies and a semiclassical picture. Another approach
that has been extensively explored is based on the observation that
in disordered conductors, spectral properties of single electron
levels (within a not-too-large energy range) are satisfactorily
described by random matrix theories \cite{Gorkov65,Efetov83}.
Remarkably enough, results of such a random matrix approach for the
a.c. absorption (hence the real part of the conductance, $g$), differ
sharply from those of the former approach. For one thing, a {\em
negative} weak field MC is predicted. To the best of our knowledge
these differences hitherto have passed practically unnoticed.

The purpose of our systematic study, reported here, is two fold. By
extending previous analyses we have attempted at bridging the
apparent differences among the various approaches alluded to above,
presenting a coherent picture of the quantum a.c. conductance and its
dependence on important parameters of the system. At the same time we
have investigated differences between the canonical and the
grand-canonical statistical ensembles, (CE) and (GCE) respectively, which
have proven to play a crucial role in the thermodynamics of such
systems \cite{Cheung88}.

	The central point of our discussion consists of the
observation that two major factors conspire to determine the behavior
of our system (where, throughout most of the present analysis the
sign of the MC is taken to be a signature of this behaviour). The
first is the dimensionless parameter $\Delta/ \gamma$, where $\Delta$
is the average level spacing at the Fermi energy and $\gamma$ is the
relaxation rate (see below) which determines the coupling of the
electronic system to the eternal degrees of freedom, hence the
broadening of the individual single electrons levels. Technically, for
$\Delta/ \gamma <1$, $(>1)$ one employs perturbative i.e. Green's
function, (non perturbative, e.g random matrix theory) techniques.
But also the physics  differs sharply between the mesoscopic
continuous spectrum limit and the microscopic discrete spectrum
limit. The second important factor is the type of the statistical ensemble
employed. We argue that even when dynamical response is concerned
different terms may contribute to the conductance depending on whether
GCE or CE is employed. Consequently in the non-perturbative limit the MC
may differ by sign and order of magnitude between the two ensembles. We also
identify such
differences, albeit small (of order $(\Delta/\gamma)^2$), in the
perturbative regime.

Our main results are summarized at the end of this paper. For the
sake of specificity we consider here quasi one--dimensional cylinders
($L_x>l$, where $L_x$ is the perimeter and $l$ -- the elastic mean
free path. The inelastic mean free path, $l_{\phi}$, is larger than
all linear scales of the cylinder). Our analysis is extendible to
other geometries (including quantum dots) where by and large, we anticipate our
qualitative predictions to hold.

Within linear response theory, we consider a small amplitude flux
component at frequency $\omega$ superimposed on a static Aharonov--Bohm flux,
$\overline\Phi$. The average conductance of a cylinder, $g$, is related to
the conductivity by a factor $S/L_x$, S being the cross section area
for the current density \cite{foot1}. Our starting point follows that
of Ref.\ \cite{Trivedi88}.  The density matrix $\hat\rho$ satisfies the
kinetic equation \begin{equation} \frac{\partial\hat\rho}{\partial
t}+i[\hat H,\hat\rho]= -\gamma(\hat\rho-\hat\rho_{QE}).
                                                            \label{keq}
\end{equation}
Here $\hat H$ is the Hamiltonian; $\gamma$ is the relaxation rate \cite{foot2}
towards an
instantaneous equilibrium state described by $\hat\rho_{QE}$
\cite{Landauer85,Trivedi88}.
A straightforward application of linear response, (cf.  Ref.\
\cite{Trivedi88}),
results in the following expression for the conductance $g$
\begin{eqnarray}
&&g(\omega,\overline\Phi)=
-\frac{1}{i\omega}\frac{\partial}{\partial\overline\Phi}
\sum_n f_n\frac{\partial\epsilon_n}{\partial\overline\Phi}+
\frac{1}{\gamma-i\omega}\frac{\partial\mu}{\partial\overline\Phi}\sum_n
\frac{\partial f_n}{\partial\epsilon}
\frac{\partial\epsilon_n}{\partial\overline\Phi}-
\nonumber \\
&&\frac{1}{\gamma-i\omega}\sum_n \frac{\partial f_n}{\partial\epsilon}
\left(\frac{\partial\epsilon_n}{\partial\overline\Phi}\right)^2+
i\sum_{n\neq m}\frac{f_n-f_m}{\epsilon_n-\epsilon_m}
\frac{|<n|\hat J|m>|^2}{\epsilon_n-\epsilon_m-\omega-i\gamma},
                                                            \label{g}
\end{eqnarray} where $\epsilon_n=\epsilon_n(\overline\Phi)$ and
$|n>=|n(\overline\Phi)>$ denote the exact single electron
eigenenergies and eigenstates; $f_n\equiv
f(\epsilon_n(\overline\Phi)-\mu(\overline\Phi))$ is the Fermi--Dirac
function \cite{foot3}, and $\hat J\equiv -\partial \hat H/\partial
\overline\Phi$ is the current operator. We shall denote the
respective impurity averaged terms on r.h.s. of Eq.\ (\ref{g}) by
$g_I, g_{II}, g_{III}$ and $g_{IV}$. We note that $g_I$ is basically
the flux derivative of the persistent current, yielding a purely
imaginary contribution to $g$, which has been evaluated both within
the GCE and CE \cite{Cheung89}. As a preliminary we note that for the
canonical ensemble $g_{II}$ cancels against $g_{III}$ at zero temperature. This
implies that in a canonical situation there is no diagonal relaxation
at zero temperature (the occupation probability of any given level is flux
independent). Hereafter we consider the frequency range
$\omega\tau\ll 1$, where $\tau$ is the elastic mean free time.

{\bf Diagrammatic approach} (valid for $\gamma\gg \Delta$).
Within the {\em grand canonical} ensemble
$\mu(\overline\Phi)=\overline\mu$ is a constant, hence $g_{II}$ vanishes. The
remaining contributions of $g_{III}$ and $g_{IV}$ yield \cite{Altshuler83}
\begin{equation}
g^{GCE}=g_0 \left[1-
\frac{1}{\pi}\frac{\Delta}{\gamma-i\omega}
F_1(\sqrt{\frac{\gamma-i\omega}{E_c}},\overline\Phi)\right],
                                                            \label{aas}
\end{equation} where $F_1(x,\overline\Phi)=\frac{1}{2}x\sinh x/(\cosh
x - \cos 4\pi\overline\Phi/\Phi_0)$ and $\displaystyle
g_0=2\frac{e^2}{\hbar}\frac{E_c}{\Delta}$ is the Drude conductance.
Here $E_c=\hbar D/L_x^2$ is the Thouless energy, $D$ is the diffusion
constant. The second term in Eq.\ (\ref{aas}), first derived in Ref.\
\cite{Altshuler83} is the $\Delta/\gamma$ weak localization correction to
the Drude result \cite{foot4}. We recall that indeed systematic
diagrammatic theory analysis may be presented as an expansion in the
small parameter $\Delta/\gamma$. The temperature, dependence of
$g^{GCE}$ comes mainly from the dependence of $\gamma$ on $T$. The
real part of the conductance, $\Re g^{GCE}$, gives rise to positive
MC at small flux (see inset to Fig. 1).

Within the {\em canonical ensemble} the number of electrons in each ring
remains unchanged as the flux is varied, which implies $\overline\Phi$
dependence of $\mu$. To evaluate the term $g_{II}$ we neglect the
$\overline\Phi$ dependence of $f$ (giving rise to higher order
terms in $\Delta/\gamma$).
We obtain at $T\ll \gamma$ \cite{foot5}
\begin{equation}
g^{CE}_{II}=g_0\frac{2}{\pi^2}\frac{\Delta}{\gamma-i\omega}
\frac{\Delta}{\gamma}
F_2(\sqrt{\frac{\gamma}{E_c}},\overline\Phi),
                                                            \label{gtwo}
\end{equation} where
$F_2(x,\overline\Phi)=x^2\frac{\partial}{\partial x}
\frac{F_1(x,0)-F_1(x,\overline\Phi)}{x}$. The temperature dependence
of $g^{CE}$ is described below. We note the correspondence between
$g^{CE}_{II}$ and the typical single level current:
$\langle i_n^2(\overline\Phi)\rangle=-4\pi(\Delta E_c/\Phi_0^2)
F_2(\sqrt{\Delta/(\pi E_c)},\overline\Phi)$ \cite{Cheung89,Gefen92}. To
evaluate the last two terms of Eq.\ (\ref{g}), $g^{CE}_{III}+g^{CE}_{IV}$
we expand $f_n$ about the flux and impurity average chemical
potential $\overline\mu$. The result can be written as \cite{foot6}:
\begin{equation} g^{CE}_{III}+g^{CE}_{IV}=g^{GCE}+
\frac{\Delta}{-i\omega}
\left[\eta(\omega,\overline\Phi)-\eta(0,\overline\Phi)\right],
                                                            \label{gthree}
\end{equation}
with $$\eta(\omega,\overline\Phi)=\frac{1}{\Delta}
\langle(\mu(\overline\Phi)-\overline\mu)
\frac{\partial}{\partial\mu}\sum_{n,m}
\frac{(f_n-f_m)|J_{nm}|^2}{\epsilon_n-\epsilon_m-\omega}\rangle.$$
Evaluation
of $\eta$ necessitated the calculation of a considerable number of diagrams
with numerous energy integration ranges.
A straightforward but lengthy calculations give
\begin{equation}
\eta(\omega,\overline\Phi)=g_0\frac{2}{\pi^2}
\left[\frac{2\Delta}{-i\omega}
\left(F_1(\sqrt{\frac{\gamma-i\omega}{E_c}},\overline\Phi)-
F_1(\sqrt{\frac{\gamma}{E_c}},\overline\Phi)\right)-
\frac{\Delta}{\gamma-i\omega}F_1(\sqrt{\frac{\gamma-i\omega}{E_c}},
\overline\Phi)\right]+[\overline\Phi\rightarrow 0].
                                                            \label{eta}
\end{equation}
Hereafter we shall
discuss the dissipative part of $g$. We note that the contributions
Eqs.\ (\ref{gtwo}) and (\ref{gthree}) to the MC are comparable
and of the same sign. Indeed $F_i(x,\overline\Phi)$ are oscillatory functions
of $\overline\Phi$ with a period $\Phi_0/2$ and an amplitude of order
unity (for $x<1$). We stress that all terms except $g^{GCE}$ (i.e. all terms
particular to the CE), contain an explicit temperature dependence.
We find that their contribution is roughly a constant $\sim
g_0(\Delta/\gamma)^2$ for $T\ll\gamma$, then decays as a power law
$\sim g_0\Delta^2/(\gamma T)$ for $\gamma\ll T\ll E_c$ and finally vanishes
exponentially $\sim g_0\Delta^2/(\gamma T)\exp(-\sqrt{T/E_c})$
for $E_c\ll T$ \cite{foot7}.

The flux dependence of the canonical terms $\Re(g^{CE}-g^{GCE})$ is depicted in
Fig. \ref{fig1} for few values of $T$ and $\gamma$. We note that the canonical
terms exhibit flux dependence opposite to the commonly accepted GCE weak
localization behavior (thus showing the tendency to the {\em negative} MC).
This dependence is eventually suppressed at temperatures  $T>E_c$.

{\bf Non--perturbative regime}. For $\gamma<\Delta$ one is not able
to employ perturbative techniques. We
thus adopt another approach following the treatment of Refs.\
\cite{Gorkov65} and \cite{Shklovskii81} of the a.c.  absorption in an
applied electric field. Certain important modifications are due.
Within the GCE the real part of the diagonal contribution
$g_{III}^{GCE}$ ($g_{II}^{GCE}\equiv 0$) is given by
$g_{III}^{GCE}=\frac{2e^2}{\pi
h}\frac{<i_n^2(\overline\Phi)>\Phi_0^2}{\Delta(\gamma -i\omega )}$.
This last equality may be written as (cf. Eq.\ (\ref{gtwo})),
\begin{equation}
	g_{III}^{GCE} = -g_0\frac{2}{\pi}\frac{\Delta}{\gamma-i\omega}
F_2(\sqrt{\frac{\Delta}{\pi E_c}},\overline\Phi),\label{gd}
	\end{equation}

To obtain Eq.(\ref{gd}) we have employed some of the results of
Ref. \cite{Cheung89,Gefen92,Atland92}.
This evidently leads to a positive MC, with  an amplitude
$2g_0\left(\frac{\Delta}{\pi\gamma}\right)\frac{\gamma^2}{\gamma^2+\omega^2}$
\cite{foot8}.
For $\omega\ll\gamma$ we recover the $1/\gamma$ divergence discussed
previously for intralevel absorption \cite{Landauer85,Trivedi88}.
Unlike in the perturbative regime, here this is the dominant contribution.

To calculate the off--diagonal contributions we first note that we
may perform separately the ensemble average  of
  $|<n|\hat J|m>|^2$ and the terms that depend on the eigenvalues
$\epsilon_n$,
 $\epsilon_m$ (see Eq.(\ref{g})). The former being averaged over an
energy interval $E_c$ is replaced by a constant, whose value is determined to
 be $g_0\Delta^2/\pi$, compatible with the requirement that $g\approx
g_0$ for $\gamma>\Delta$
\cite{Gorkov65,foot9,Efetov93}.
 We thus may write $g_{IV}$ as

\begin{equation}
g_{IV}^{GCE}=\frac{g_0}{i\pi}\int_{-\infty}^{\infty}
\frac{d\epsilon}{\epsilon-\omega-i\gamma} \left[\frac{1}{\Delta^2}
\langle\sum_{n\neq m}\delta(\epsilon-\epsilon_n+\epsilon_m)
\frac{f_n-f_m}{\epsilon_m-\epsilon_n}\rangle_{GCE}\right]
                                                            \label{goff}
\end{equation}
where $\langle\ldots\rangle_{GCE}$ refers to averaging under grand canonical
conditions (e.g. assuming $\mu$ to be uniformly distributed over an energy
interval much larger than $\Delta$). The quantity in the brackets
is $R(\epsilon,\overline\Phi)$ -- the level pair correlation function.
It may be shown that , to leading order in $x\equiv\pi\epsilon/\Delta$
\cite{Atlandun,Montambaux}
\begin{equation}
R(\epsilon,\overline\Phi)=\left\{ \begin{array}{ll}
\frac{\pi}{6}x,  & \nu\ll x\ll 1; \\
\frac{\sqrt{\pi}}{6}\frac{x^2}{\nu},  \hskip .75cm &  x\ll \nu\ll 1;\\
\frac{x^2}{3},     & x\ll 1\ll\nu;
\end{array}
\right.
                                                            \label{R}
\end{equation}
with $\nu\equiv (2\pi)^{3/2}\sqrt{E_c/\Delta}\, (\overline\Phi/\Phi_0)$.
The leading
contribution to $g_{IV}^{GCE}$ are respectively ($x$ to be replaced by
$z\equiv\pi(\omega+i\gamma)/\Delta$ \cite{foot11}):
\mbox{$g_0z(\pi+2i(\ln z+\Gamma-2))/6$} in orthogonal;
$g_0z(2i\ln \nu + \sqrt{\pi}z/\nu)/6$ in intermediate;
and  $g_0z(z-2i)/3$ in unitary cases,
where $\Gamma$ is Euler's constant.
This off--diagonal term is marked  by a negative MC of amplitude
$g_0(\gamma/\Delta)\ln (\gamma/\Delta)$ (at $\omega=0$).
We note, though, that this
off-diagonal term is only a small correction on top of the diagonal
contribution, (the latter being of order of $\Delta/\gamma$ when $\omega=0$,
cf Eq.\ (\ref{gd})). The total MC (including the dominant $\Delta/\gamma$
diagonal contribution Eq.\ (\ref{gd})) gives rise to a positive MC.
There is practically no explicit $T$ dependence in the GCE case.

The situation is markedly different for the CE. As was noted above,
the two diagonal contributions to  $g^{CE}$ are offset at $T=0$
($g^{CE}_{diag}\equiv g^{CE}_{II}+g^{CE}_{III}=0$, for $T=0$).
This offset
is highly temperature dependent. The difference
$|g^{CE}_{diag}-g^{GCE}_{diag}|$ decreases significantly at $T\approx\Delta$,
and continues to decrease further as $T^{-1}$ until at $T\ge E_c$
differences between two ensembles disappear. Thus the diagonal contribution,
giving rise to positive MC,  dominates at sufficiently high $T$
see Fig.\ \ref{fig2}.

As for the off--diagonal contribution, the average in
Eq.\ (\ref{goff}) is to be
replaced by $\langle\ldots\rangle_{CE}$, implying that particle number,
rather than chemical potential $\mu$, is a uniformly distributed
random parameter. We note that
\mbox{$\langle\ldots\rangle_{CE}\approx\langle\ldots\rangle_{GCE}
\frac{(n-m)\Delta}{\epsilon}$}.
Making further the approximation
$(n-m)\Delta\approx \Delta\mbox{sign}\epsilon+ \epsilon$, we obtain
(cf.  Eq.\ (\ref{goff}))
\begin{equation}
g^{CE}_{IV}=g^{GCE}_{IV}+\frac{g_0}{i\pi}
\int_{-\infty}^{\infty}d\epsilon
\frac{R(\epsilon,\overline\Phi)}{\epsilon-\omega-i\gamma}
\frac{\Delta}{|\epsilon|}
                                                      \label{gfour}
\end{equation} Note the extra $|\epsilon|^{-1}$ factor in Eq.(\ref{gfour}).
Consequently, the sensitivity to small magnetic flux (arising due
to  change in a level statistic), is more pronounced in the CE than in
the GCE. We obtain for the r.h.s. of  Eq.(\ref{gfour})
$g_0\pi^2/6$ in orthogonal;
\mbox{$g_0\sqrt{\pi}(\pi z+2 i\ln z)/(6\nu)$} in intermediate; and
$g_0(\pi z+2 i\ln z)/3$ in unitary regimes. At
$T=0, \omega=0$ this yields for the real part of conductance, $g_0\pi^2/6$
for orthogonal and $g_0(2\pi/3)(\gamma/\Delta)\ln(\Delta/\gamma)$  for
unitary  cases.
At $T=0, \gamma=0$ we obtain $g_0\pi^2/6$
and $g_0(\pi^2/3)\omega$ respectively \cite{foot13}.
This leads to a large negative MC of amplitude
$\approx g_0$ (to be compared with the small corresponding
off-diagonal contribution in the GCE case). Numerical results for the
flux dependence of $g^{CE}$ are shown in Fig.\ref{fig2}. Our
numerical data are compatible with our analytical results, showing
for the CE a transition from negative to positive MC as $T$ increased.
More details of our study will be published elsewhere.

In summary, we have reported here a comprehensive study of the
conductance of mesoscopic rings in various regimes of the relevant
parameters. In particular we have noted that: (1) Within the CE the
MC may change sign as function of the parameter $\Delta/\gamma$,
going from the discrete level to the continuum limit. (2) The
incipient transition may be seen from a perturbation expansion
 in powers of $(\Delta/\gamma)$. The $(\Delta/\gamma)^2$
term possesses a sign opposite to the $(\Delta/\gamma)$ term and it
eventually takes over as $(\Delta/\gamma)$ approaches unity. (3) The
diagonal term with the CE is highly temperature sensitive. In the
$\Delta/\gamma>1$ regime this gives rise to a sign change of the MC
as a function of temperature. Within the GCE the corresponding term
is by far more robust to temperature, rendering the MC always
positive. (4) Differences between the two ensembles, accentuated at
low temperatures, diminish as a power law with temperature up to
$T\approx E_c$. (5) We have found contributions to the conductance
which mainly depend on elastic scattering (e.g Eq.\ (\ref{gfour})),
while others are dramatically dependent on the inelastic rate (e.g.
the diagonal contribution Eq.\ (\ref{gd})). This sheds light on
previous arguments concerning the relative importance of elastic and
inelastic processes in determining $g$ \cite{Landauer85,Imry86}. We finally
note that within present technology, one is not too far from
achieving the limit $\gamma<\Delta$ in experiment \cite{sivan}.

We have benefited from useful discussions with G. Montambaux.
We also acknowledge comments by M.T. Beal--Monod,
O. Entin, B. Shklovskii, B.Spivak. A.K. and Y.G. acknowledge the hospitality
of H. Bouchiat and G. Montambaux at Orsay. This work was supported by the
German--Israel Foundation (GIF) and the U.S.--Israel Binational
Science Foundation (BSF).

\figure{Flux dependence of the canonical terms
$\Re(g^{CE}-g^{GCE})$ (in units of $e^2/\hbar$) for
$E_c/\Delta=20, \omega=0$. (Solid line) $T=0,
\gamma=5\Delta$; (dashed line) $T=0, \gamma=10\Delta$; (doted line)
$T=10\Delta, \gamma=5\Delta$. Inset: the GCE conductance
(solid line) $\gamma=5\Delta$; (dashed line) $\gamma=10\Delta$.
\label.\label{fig1}}

\figure{Flux dependence of the canonical average conductance
(in units of $e^2/\hbar$), obtained from numerical simulations on the
Anderson model for a 3D ring of dimensions $30\times 4\times 4$. Here
$E_c/\Delta=7/2\pi, \omega=0$, $\gamma=.3\Delta$. The numerical
results are averaged over energy range and over two disorder
configurations. The data obtained for the CE at $T=E_c$ are
identical with the GCE.  \label{fig2}}


\begin{references}

\bibitem{Landauer85}R. Landauer and M. Buttiker, Phys. Rev. Lett. {\bf
54}, 2049 (1985).

\bibitem{Imry86}Y. Imry, and N. S. Shiren, Phys. Rev. {\bf B 33}, 7992
(1986);  Y. Gefen and O. Entin-Wohlman, Ann.Phys.
{\bf 206}, 68 (1991).

\bibitem{Trivedi88}N. Trivedi, and D. Browne, Phys. Rev. {\bf B 38}, 9587
(1988).

\bibitem{Landauer92}R. Landauer, Physica Scripta {\bf 42}, 110
(1992).See also, R. A. Serota, J. Yu and Y. Kim, Phys. Rev. {\bf B 42}, 9724
(1990).

\bibitem{Altshuler83}B. L. Altshuler, A. G. Aronov, and B. Z. Spivak,
Pis'ma Zh. Eksp. Teor. Fiz. {\bf 33}, 101 (1981)
[JETP Lett. {\bf 33}, 94 (1981)].

\bibitem{Gorkov65}L. P. Gorkov, and G. M. Eliashberg, Zh. Eksp. Teor.
Fiz. {\bf 48}, 1407 (1965) [Sov. Phys. JETP {\bf 21}, 940 (1965)].

\bibitem{Efetov83}K. B. Efetov, Adv. Phys. {\bf 32}, 53 (1983).

\bibitem{Cheung88}H. F. Cheung, Y. Gefen, E. K. Riedel, and W. H. Shih,
Phys. Rev. {\bf B 37}, 6050 (1988); H. Bouchiat, and G. Montambaux,
J. Phys. (Paris) {\bf 50}, 2695 (1989).

\bibitem{foot1}The advantage of considering this geometry is twofold. The
induced electromotive force around the ring is hardly screened, rendering
differences between external and internal fields negligible. Also canonical
 conditions (particle conservation) are readily obtained.

\bibitem{foot2}Repeating our calculation with two different relaxation rates,
diagonal and off-diagonal, resulted in no qualitative changes from the
present analysis. The $\gamma$ that appears in the last term of
Eq.\ (\ref{g}) is in fact the off--diagonal rate $\gamma_{off}$.

\bibitem{foot3}Within the CE the correct distribution differs from the
Fermi--Dirac function. For the quantities studied here this may give rise to
quantitative corrections of at most 15\% (A. Kamenev and Y. Gefen unpublished),
but will not modify the qualitative picture derived here.

\bibitem{Cheung89}H. F. Cheung, E. K. Riedel, and Y. Gefen,  Phys. Rev. Lett.
{\bf 62}, 587 (1989); A. Schmid,  Phys. Rev. Lett. {\bf 66}, 80 (1991);
B. L. Altshuler, Y. Gefen, and Y. Imry, Phys. Rev. Lett. {\bf 66}, 88 (1991).

\bibitem{foot4}Terms of the order of $(\Delta/\gamma)^2$ have been ignored
here. We note, though, that precisely the same terms contribute also to the
canonical conductance. Thus, as far as the difference $g^{GCE}-g^{CE}$ is
concerned our results are valid to order $(\Delta/\gamma)^2$.

\bibitem{foot5}In fact the inelastic rate appearing under the integral is
$\gamma_{out}$, the quasi inelastic rate associated with the impurity lines,
while the one in front of it is the diagonal relaxation, $\gamma_{diag}$.
Throughout this paper we ignore differences among the \mbox{various $\gamma$}.

\bibitem{Gefen92}Y. Gefen, B. Reulet, and H. Bouchiat, Phys. Rev.
{\bf B 46}, 15922 (1992); A. Kamenev, Y. Gefen, Phys. Rev. Lett. {\bf 70},
1976 (1993);

\bibitem{foot6}To pursue diagrammatic calculation, we have assumed that
$\gamma_{off}$ is negligible compared with $\gamma_{out}$, which arises from
the ladder diagrams. We have explicitly checked that the inelastic rate
entering $g^{GCE}$ is, in fact, the sum $\gamma_{\phi}+\gamma_{off}$,
$\gamma_{\phi}$ being the dephasing rate.

\bibitem{foot7}A. Kamenev; Y. Gefen  unpublished.

\bibitem{Shklovskii81}B. I. Shklovskii, Pis'ma Zh. Eksp. Teor.
Fiz. {\bf 36}, 287 (1982) [JETP Lett. {\bf 36}, 352 (1982)].

\bibitem{Atland92}
A.~Altland, S.~Iida, A.~M\"uller-Groeling, and H.~A. Weidenm\"uller,
Europhys. Lett. {\bf 20}, 155 (1992);

\bibitem{foot8}Here $\gamma=\gamma_{diag}$ is the diagonal relaxation rate.

\bibitem{foot9}Note the difference with Ref.\ \cite{Gorkov65}, where due to
different boundary conditions it was assumed that the matrix elements of
$\hat X$ are constant.

\bibitem{Efetov93} K. B. Efetov and S. Iida, Phys. Rev. B {\bf 47},
15794 (1993).

\bibitem{Atlandun} A. Altland, K.B. Efetov and S. Iida, J. Phys. {\bf
A26}, 3545 (1993).

\bibitem{Montambaux}N. Dupuis, and G. Montambaux, Phys. Rev. {\bf B 43},
14390 (1991).

\bibitem{foot11}Here $\gamma=\gamma_{off}$.

\bibitem{foot13} We stress, though, that within our analysis $\gamma$
is taken to
be a constant independent of energy. There are, however, important
circumstances where $\gamma$  decreases to  zero  with the relevant
energy scale, which renders the limit $\omega=0$, $\gamma>0$ invalid.
The other limit $\omega>0$, $\gamma=0$ is meaningful even under
these circumstances. We thank B.Spivak for useful discussions
concerning this point.

\bibitem{sivan}
U. Sivan, F. P. Milliken, K. Milkove, S. Rishton, Y. Lee, J. M. Hong,
V. Boegli, D. Kern and M. deFranza, preprint (1993).

\end{references}
\end{document}